\documentclass[sigconf]{acmart}

\usepackage{censor}
\usepackage{subcaption}
\usepackage{lipsum}
\usepackage{graphicx}
\usepackage{xspace}
\usepackage{amsmath}

\def\BibTeX{{\rm B\kern-.05em{\sc i\kern-.025em b}\kern-.08emT\kern-.1667em\lower.7ex\hbox{E}\kern-.125emX}}
    
\copyrightyear{2018}
\acmYear{2018}
\setcopyright{none}
\acmConference[RecSys '19]{RecSys '19: The ACM Conference Series on Recommender Systems}{September 16--20, 2019}{Copenhagen}
\acmBooktitle{RecSys '19: The ACM Conference Series on Recommender Systems, September 16--20, 2019, Copenhagen, Denmark}
\acmPrice{15.00}
\acmDOI{10.1145/1122445.1122456}
\acmISBN{978-1-4503-9999-9/18/06}

\renewcommand\footnotetextcopyrightpermission[1]{} 
\begin{document}

%
\title{An Explainable Autoencoder For Collaborative Filtering Recommendation}

%

\author{Pegah Sagheb Haghighi}
\affiliation{%
  \institution{Knowledge Discovery and Web Mining Lab, Dept. of Computer Science and Engineering, University of Louisville}
  \streetaddress{Louisville, KY 40292}
  \city{Louisville}}
\email{pegah.saghebhaghighi@louisville.edu}

\author{Olurotimi Seton}
\affiliation{%
  \institution{Knowledge Discovery and Web Mining Lab, Dept. of Computer Science and Engineering, University of Louisville}
  \streetaddress{Louisville, KY 40292}
  \city{Louisville}}
\email{rotimi.seton@louisville.edu}

\author{Olfa Nasraoui}
\affiliation{%
  \institution{Knowledge Discovery and Web Mining Lab, Dept. of Computer Science and Engineering, University of Louisville}
  \streetaddress{Louisville, KY 40292}
  \city{Louisville}}
\email{olfa.nasraoui@louisville.edu}

%

\renewcommand{\shortauthors}{Damak and Nasraoui}

%
\begin{abstract}

Autoencoders are a common building block of Deep Learning architectures, where they are mainly used for representation learning. They have also been successfully used in Collaborative Filtering (CF) recommender systems to predict missing ratings. Unfortunately, like all black box machine learning models, they are unable to explain their outputs. Hence, while predictions from an Autoencoder-based recommender system might be accurate, it might not be clear to the user why a recommendation was generated. In this work, we design an explainable recommendation system using an Autoencoder model whose predictions  can be explained using the neighborhood based explanation style. Our preliminary work can be considered to be the first step towards an explainable deep learning architecture based on Autoencoders.

\end{abstract}

%
%

\settopmatter{printacmref=false}

%
\keywords{recommender systems, explainability, neural networks, autoencoder, deep learning}

%
%
\maketitle

\section{Introduction}

The use of information filtering tools that discover or suggest relevant and personalized items has become essential to avoid information overload. For instance, recommender systems assist users by providing them with personalized suggestions. They recommend items in a variety of domains (music, movies, books, travel recommendation, etc). However, the most accurate recommendation models tend to be black boxes \cite{Herlocker01} that cannot justify why items are recommended to a user. The ability of a recommender system to explain the reasoning of its recommendation can serve as a bridge between humans and recommender systems. Explanations can serve different purposes such as building trust, improving transparency and user satisfaction and helping users make informed decisions \cite{Herlocker01, Tintarev02}.
Latent factor models have been the state of the art in Collaborative Filtering recommender systems. For instance, Matrix Factorization (MF) learns hidden interactions between entities to predict possible user ratings for items.  However, a common challenge with this method is the difficulty of explaining recommendations to users using the latent dimensions. To solve this problem, some explainable recommendation algorithms have recently been proposed. For instance, Explicit factor models (EFM) \cite{Zhang2014} generate explanations based on the explicit features extracted from users' reviews. Explainable Matrix Factorization (EMF) \cite{Abdollahi2016, Abdollahi2017} is another algorithm  that uses an explainability regularizer or soft constraint in the objective function of classical matrix factorization. The constraining term tries to bring the user and explainable item's latent factor vectors closer, thus favoring the appearance of explainable items at the top of the recommendation list. When it comes to Autoencoder recommender models however, very little research has been conducted to address the explainability of the predictions. In this paper, we propose an Autoencoder model that finds top \textit{n} accurate recommendations that are explainable without the use of additional data modalities such as content. The rest of the paper is organized as follows: Section 2 reviews related work, Section 3 presents a novel explainable Autoencoder for Collaborative Filtering. Section 4 presents the experimental results. Finally, Section 5 concludes the paper. For clarity, all notations used in this paper are summarized in Table~\ref{table:T1}.

\section{Related Work}
\begin{figure}[h] 
\begin{center}  
\includegraphics[scale=0.50]{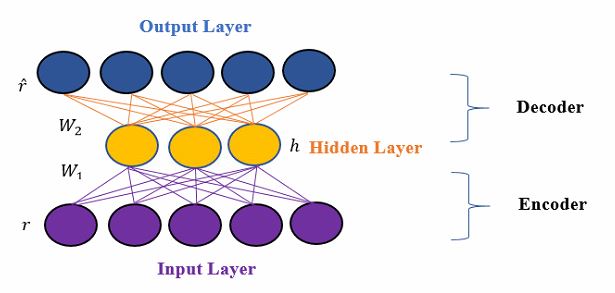}
\caption{\small \sl An Autoencoder Architecture}
\label{fig:1}
\end{center}
\end{figure}

Recently, a growing body of research has involved using neural networks such as Autoencoders for collaborative filtering. An Autoencoder is an unsupervised learning model and a form of neural network which aims to learn important features of the dataset. Architecturally, an Autoencoder is a three-layer feed-forward neural network consisting of an input layer, a hidden layer and an output layer. Figure~\ref{fig:1} depicts a classical Autoencoder architecture. An Autoencoder consists of two parts, an encoder and a decoder. The encoder maps input features into a hidden representation and the decoder aims to reconstruct the input features from the hidden features \cite{Hinton2006}. The encoder and decoder may each have a deep architecture which consists of multiple hidden layers.  It is worth noting that the output layer has the same dimension or number of neurons as the input layer. 

\begin{table*}[t]
\caption{Summary of notations} 
\centering 
\begin{tabular}{c c|c c} 
\hline\hline 
Symbol & Description & Symbol & Description \\ 
\hline 
$\mathcal{U}$ & set of all users & $U_{i,x}$ & set of users given rating $x$ to item $i$ \\ 
$\mathcal{I}$ & set of all items & $\mathcal{R}^{test}$ & ratings in the test set \\
$\sigma$  & sigmoid activation functions & $\sigma^{\prime}$ & identity activation function \\
$\boldsymbol{b}$, $\boldsymbol{b}^\prime$  & biases & $\boldsymbol{E}$ & explainability matrix \\
$\boldsymbol{R}$ & rating matrix & $\boldsymbol{e}$ & explainability vector\\
$\boldsymbol{r}^u$ & user sparse rating vector & $N_u$ & set of users who are most similar to user $u$ \\
$\boldsymbol{r}^i$ & item sparse rating vector &  $\boldsymbol{W}_1$, $\boldsymbol{W}_2$ & connection weights between layers \\[1ex] 
\hline 

\end{tabular}
\label{table:T1} 
\end{table*}

A collaborative filtering based Autoencoder was introduced by Sedhain et al. \cite{sedhain2015autorec} and Ouyang et al. \cite{Ouyang2014}. They achieved a better accuracy than current state-of-the-art methods such as Matrix Factorization to predict missing values of the user-item matrix. Their model worked as follows: Given a hidden layer, the encoder section of an Autoencoder takes the ratings data $\boldsymbol{r}$ of each user and maps it to a latent representation $\boldsymbol{h}$, see Eq.(1). User preferences are encoded in a sparse matrix of ratings $\boldsymbol{R}$ $\epsilon$  $\mathbb{R}^{m \times n}$, where $m$ and $n$ are the number of users and items, respectively. $\mathcal{U} = \{ 1, .., m \}$ represents the set of all users and $\mathcal{I} = \{ 1, .., n \}$ represents the set of all items. Each user $u$ $\epsilon$  $\mathcal{U}$ is represented by a sparse vector $\boldsymbol{r}^{(u)} = \{ R_{u1},...,R_{un} \}$, and each item $i$ $\epsilon$  $\mathcal{I}$ is represented by a sparse vector $\boldsymbol{r}^{(i)} = \{ R_{1i},...,R_{mi} \}$. The hidden or latent representation is given by

\begin{equation}
    \boldsymbol{h} = \sigma(\boldsymbol{W}_{1}. \boldsymbol{r} + \boldsymbol{b}) 
\end{equation}

where $\sigma$ is the activation function - in this case the sigmoid function ($\frac{1}{1+ e^{-x}}$), $\boldsymbol{W}_1$ is a weight matrix, and $\boldsymbol{b}$ is a bias vector.

The decoder maps the latent representation $\boldsymbol{h}$ into a reconstruction output $\boldsymbol{\hat{r}}$ given by

\begin{equation}
\boldsymbol{\hat{r}} = \sigma^{\prime}(\boldsymbol{W}_2. \boldsymbol{h} + \boldsymbol{b}^\prime) 
\end{equation}

where $\sigma^{\prime}$ is an activation function - in this case Identity.
The goal of the Autoencoder is to minimize the reconstruction error by minimizing a reconstruction loss function as follows

\begin{equation}
{\mathrm{min}} \sum_u \parallel \boldsymbol{r}^u  -  \boldsymbol{\hat{r}}^u \parallel^2  + \frac{\lambda}{2} . (\parallel \boldsymbol{W}_1 \parallel_F^2 + \parallel \boldsymbol{W}_2 \parallel_F ^2) 
\end{equation}

where $\lambda$ is the regularization term coefficient and $||.||$ is the Frobenius norm.\\

Wu et al. \cite{Wu2016} introduced the Collaborative Denoising Auto-Encoder (CDAE) which has one hidden layer that encodes a latent vector for the user. Strub and Mary \cite{Strub2015} proposed a Stacked Denoising AutoEncoders neural network (SDAE) with sparse inputs for recommender systems. \cite{Zhang2017} developed an Autoencoder model which learns feature representations from side information to improve recommendations.

When it comes to neural network models, there are a few related works that focus on explainable recommender systems. Seo et al. \cite{seo2017interpretable} designed an interpretable Convolutional Neural Network (CNN) model that predicts ratings using text reviews and ratings data.  The authors used two attention layers: a local attention layer and a global attention layer. The local layer learns which keywords are more informative, whereas the global layer learns word sequences from the original review.  Finally, the outputs of these two layers are combined and passed through the fully connected layers.

Costa et al. \cite{costa2018automatic} presented a model based on Long Short Term Memory (LSTM) Recurrent Neural Network (RNN) to automatically generate natural language explanations for recommender systems. The authors tried to improve the work proposed by \cite{lipton2015capturing} by considering a vector of auxiliary data instead of only one dimension of auxiliary information.  The vector of auxiliary data in their model is a set of rating scores for different features of items.

Chen  et  al. \cite{chen2018visually}  used  both  implicit  feedback  and  textual  reviews  to  introduce a visually explainable recommendation. By combining images and reviews, their model generated natural language explanations, which could describe the highlighted image regions to the user.

Recently, Bellini et al. \cite{bellini2017auto} introduced the SEM-AUTO approach which makes use of Autoencoders and semantic information via the DBpedia knowledge graph. The authors mapped a set of categorical features of an item into the hidden layers for each user. As a result, this model is capable of retrieving the categorical information of the items rated by the user. The proposed model aims to capture the shared side information of all the items rated by a user.  Hence, the information associated with positively rated items get a high value (close to 1), and for those with negatively rated items will tend to be closer to 0.  This approach can therefore help Autoencoders find the potential features that match user’s tastes.

To summarize our review, neural network models and in particular Autoencoders, have recently received increasing attention in the Recommender Systems field. However, like other latent factor models, Autoencoders are black-box models that are unable to provide an explanation for their output. The research questions that we are trying to answer are: (1) Can we generate an explainable Collaborative Filtering recommender system using the Autoencoder architecture? (2) Can the explainable Autoencoder maintain an accuracy that is comparable to the classical Autoencoder?

\section{The Explainable Autoencoder (E-AutoRec)}

\begin{figure}[h] 
\begin{center}  
\includegraphics[scale=0.50]{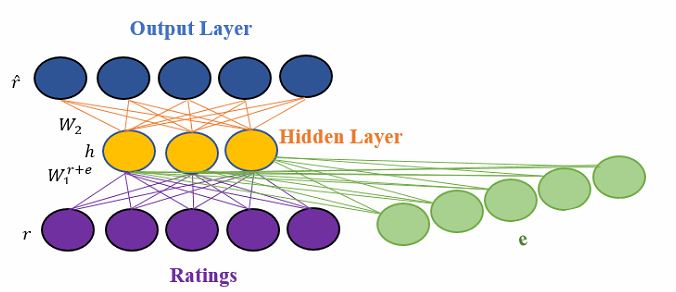}
\caption{\small \sl Explainable AutoEncoder (E-AutoRec)}
\label{fig:2}
\end{center}
\end{figure}

Our work is inspired by U-AutoRec \cite{sedhain2015autorec} and \cite{Zhang2017} with one important distinction where we feed an explainability vector of the same size as the ratings to additional input layer units in a way that is similar to adding an explainability layer to the Explainable Restricted Boltzmann Machine \cite{abdollahi2016explainable}. The architecture of our model is shown in Figure~\ref{fig:2}. In \cite{Strub2015}, the rationale for adding side information to the Autoencoder inputs in the form of an additional (hybrid) modality such as user profile features, was that when many input ratings are missing, the side information would take over to solve the cold start problem or to alleviate the sparsity of the ratings. In our case, the side information consists of the explainability scores, which likewise can help compensate for the sparsity of the ratings, but unlike the hybrid side information in \cite{Strub2015}, our side information makes up for lack of rating data using the explainability of an item to a user, which is itself a byproduct of the ratings only and not external or hybrid data. In fact the explainability of an item to a user is derived directly from the rationale of collaborative filtering by aggregating the nearest neighboring users’ ratings on that item. In the following, the reader should refer to Table~\ref{table:T1} as a reference for all the notation symbols used in our equations.

The proposed encoder takes as input each user’s ratings vector $\boldsymbol{r} \epsilon \hspace{0.1cm} \mathbb{R}^{n}$ and a user’s explainability score vector $\boldsymbol{e} \hspace{0.1cm} \epsilon \hspace{0.1cm} \mathbb{R}^{n}$ that is pre-computed offline for each user (calculated as explained in Eq.(6)-(8), similarly to the explainability scores in \cite{Abdollahi2017}) and maps them to the latent feature vector $\boldsymbol{h} \epsilon \hspace{0.1cm} \mathbb{R}^{k}$, where  $k$ is the number of hidden units, as follows

\begin{equation}
\boldsymbol{h} = \sigma(\boldsymbol{W}_1^{r+e}. (\boldsymbol{r+e}) + \boldsymbol{b}) 
\end{equation}

While both the output, see Eq. (2), and the loss function, see Eq. (3), remain the same. It is worth noting that $\boldsymbol{W_{1}^{r+e}} \epsilon \hspace{0.1cm} \mathbb{R}^{2 \times n \times k}$ is the connection weight matrix between the hidden layer and the input layer accepting both the ratings and the explainability scores.\\

We computed the explanation scores offline based on a collaborative filtering neighborhood rationale. The nearest neighbors are determined based on the cosine similarity. The explainability score for the user-based neighbor explanation \cite{Abdollahi2017} is given by:
\begin{equation}
    Exp. Score_{(u,i)} =  \boldsymbol{E}(r_{u,i}|N_u) = \sum_{x \epsilon X} x \times Pr(r_{u,i} = x | u \hspace{0.1cm} \epsilon \hspace{0.1cm} N_{u})
\end{equation}

where

\begin{equation}
    Pr(r_{u,i} = x| v \hspace{0.1cm} \epsilon \hspace{0.1cm}  N_u) = \frac{|N_u  \cap U_{i,x}|}{|N_u|}
\end{equation}

$r_{u,i}$ is the rating user $u$ gave to item $i$, $U_{i,x}$ is the set of users who have given the same rating $x$ to item $i$, and $N_u$ is the set of neighbors for user $u$.

The explainability matrix, $\boldsymbol{E}$, is a thresholded version of the neighborhood explanation scores \cite{Abdollahi2017}, as follows

\begin{equation}
  E_{u,i} =
    \begin{cases}
      Exp. Score_{(u,i)} & \text{if $Exp. Score_{(u,i)}$ $>$ $\theta$}\\
      0 & \text{otherwise}
    \end{cases}       
\end{equation}      

where $\theta$ is a user-defined threshold value to accept if item $i$ is explainable to user $u$. $Exp. Score_{(u,i)}$ is the expected value calculated in Eq. (5).

\section{Experimental Results}
 \subsection{Data and Metrics}
\begin{figure}[h] 
\begin{center}  
\includegraphics[scale=0.40]{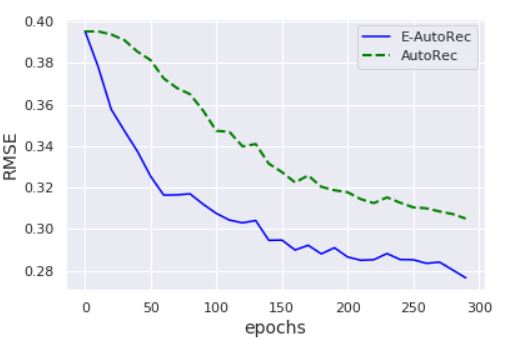}
\caption{\small \sl RMSE vs epochs on test-set}
\label{fig:3}
\end{center}
\end{figure}

We tested our approach on the MovieLens benchmark data. This dataset consists of 100K ratings on a scale of 1-5, given by 943 users to 1682 movies, where each user has rated at least 20 movies. The rating data was randomly split into $90\%$ training and $10\%$ testing subsets and the ratings were normalized between 0 and 1. We compared our results with the user-based AutoRec baseline method \cite{sedhain2015autorec} which uses the same input as our Autoencoder (that is, ratings). We did not compare to \cite{Strub2015} and \cite{Zhang2017} because they are hybrid approaches benefiting from an additional user feature modality, unlike our pure rating-based collaborative filtering approach. To assess the rating prediction accuracy, we use the Root Mean Squared Error $(RMSE)$, see Eq. (8), for both the AutoRec baseline and the proposed E-AutoRec (Explainable Autoencoder) (as shown in Figure~\ref{fig:3}) and the Mean Average Precision (MAP) score Eg. (10). We also measured the explainability of the models using the Mean Explainable Precision (MEP) given in Eq. (9), as proposed by \cite{Abdollahi2016}.

\begin{equation}
    RMSE = \sqrt{\frac{1}{|\mathcal{R}^{test}|} \sum (r_{u,i} - \hat{r}_{u,i})^2}
\end{equation}

where $|R^{test}|$ is the total number of ratings in the test set, $r_{u,i}$ is the true rating and $\hat{r}_{u,i}$ is the predicted rating for user $u$ and item $i$.

The Mean Explainability Precision $(MEP)$ for the top-n recommended list $\boldsymbol{I}_{rec}^{u}$ is computed as follows: let $\boldsymbol{I}_{rec}^{u}$ be the top-n recommended list for user $u$ and $\boldsymbol{E}_{rec}^{u}$ be the set of explainable items.

\begin{equation}
    MEP @ n= \frac{1}{|\mathcal{U}|} \sum_{u \in \mathcal{U}}  \frac{|\mathcal{I}_{exp} \cap \mathcal{I}_{rec}|}{|\mathcal{I}_{rec}|}
\end{equation}

where $\mathcal{I}_{exp}$ is the set of explainable items for user $u$, $\mathcal{I}_{rec}$ is the set of items in the top-n recommendation list for user $u$, $|\mathcal{I}_{exp} \cap \mathcal{R}_{rec}|$ is the number of explainable items present in the top-n recommendation list for user $u$, and $|\mathcal{U}|$ is the number of users in the test-set.

Finally, to evaluate the top-n recommendation performance, we compute the Mean Average Precision $(MAP)$ shown in Figure~\ref{fig:4}.

\subsection{Analysis of Results}
\begin{figure}[h]
\centering
   \begin{subfigure}{0.21\textwidth}
   \centering
   \includegraphics[width=\linewidth]{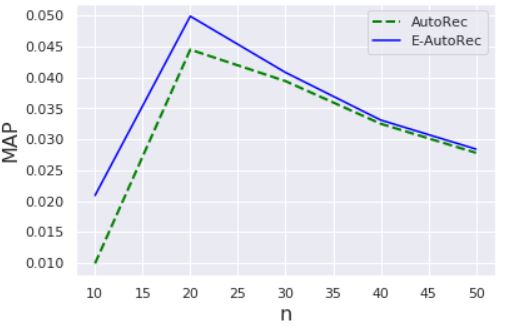}
   \caption{}
   \label{fig:4(a)} 
\end{subfigure}\quad
\begin{subfigure}{0.21\textwidth}
   \centering
   \includegraphics[width=\linewidth]{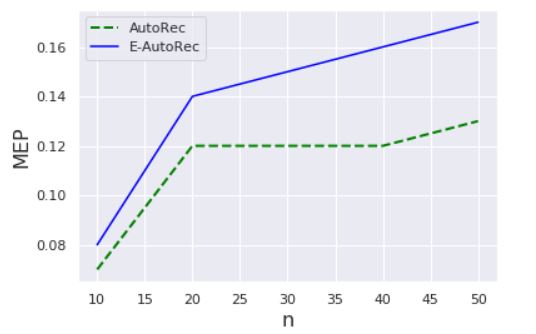}
   \caption{}
   \label{fig:4(b)}
\end{subfigure}

\centering
\caption{(a) MAP and (b) MEP vs. $n$, the size of the top-n recommendation list}
\label{fig:4}
\end{figure}

\begin{figure}[H]
\centering
   \begin{subfigure}{0.21\textwidth}
   \centering
   \includegraphics[width=\linewidth]{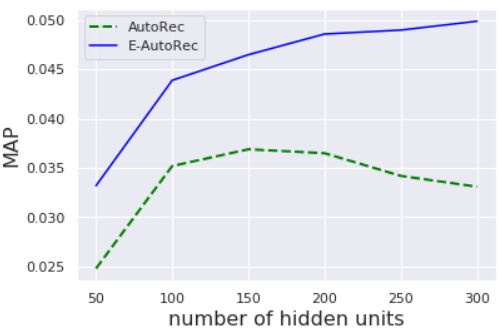}
   \caption{}
   \label{fig:5(a)} 
\end{subfigure}\quad
\begin{subfigure}{0.21\textwidth}
   \centering
   \includegraphics[width=\linewidth]{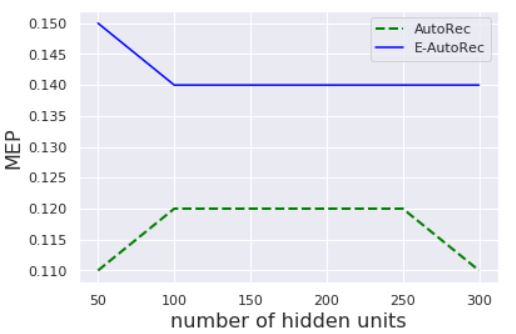}
   \caption{}
   \label{fig:5(b)}
\end{subfigure}
\centering
\caption{(a) MAP and (b) MEP vs. number of hidden units}
\label{fig:5}
\end{figure}

\begin{figure}[H]
\centering
   \begin{subfigure}{0.21\textwidth}
   \centering
   \includegraphics[width=\linewidth]{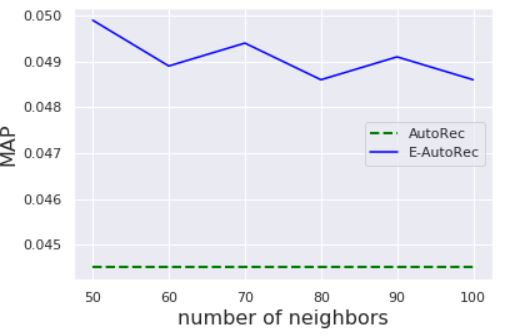}
   \caption{}
   \label{fig:6(a)} 
\end{subfigure}\quad
\begin{subfigure}{0.21\textwidth}
   \centering
   \includegraphics[width=\linewidth]{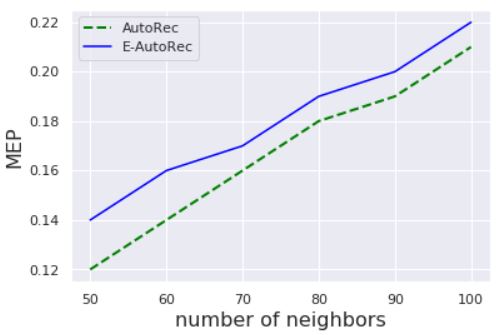}
   \caption{}
   \label{fig:6(b)}
\end{subfigure}
\begin{subfigure}{0.21\textwidth}
   \centering
   \includegraphics[width=\linewidth]{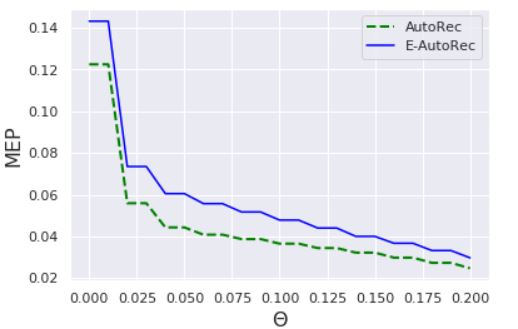}
   \caption{}
   \label{fig:6(c)}
\end{subfigure}
\centering
\caption{(a)  MAP and (b) MEP vs. number of neighbors used to compute the explainability scores (c) MEP vs explainability threshold $\theta$}
\label{fig:6}
\end{figure}

The results in Figure~\ref{fig:3}-Figure~\ref{fig:6}, show that the proposed Explainable Autoencoder (E-AutoRec) succeeds to make predictions that are at least as accurate as  the baseline Autoencoder (AutoRec), while promoting the explainability of the items in the top-n recommendation list (MEP is higher across all tested parameter ranges).

Figure~\ref{fig:4} shows the metrics MAP (a) and MEP (b) for both models, as the number of recommended items $(n)$ increases. In all experiments where we varied a parameter, we fixed the remaining parameters as follows: The explainability threshold $\theta=0$, the neighborhood size  $|N_u|=50$ and the number of hidden units $k=300$. In the case of top-n recommendations, both models appear to perform well up to a top 20 recommendation list, beyond which both suffer a significant drop in performance (Figure~\ref{fig:4(a)}). However, the MEP for both models improves well beyond the top 20 recommended items (Figure~\ref{fig:4(b)}).

To study the effect of the number of hidden units and neighborhood size (used for generating the Explainability Matrix(E)), we varied both values and measured the resulting MAP and MEP metrics, as shown in Figure~\ref{fig:5} and Figure~\ref{fig:6}. These results confirm that the accuracy and explainability are significantly better in E-AutoRec, compared to the baseline AutoRec, across all parameter ranges. Although at first, one’s intuition is that accuracy has to pay a price for each increase in explainability, this does not occur in E-AutoRec. This is because the explainability values are integrated within the actual  machine learning process, and the learning process is based on the same reconstruction loss function for rating prediction, thus maintaining the prediction accuracy. However explainability is enhanced because the explainabilty scores are used as side information, thus  compensating for sparse inputs, and as a result effectively enhancing both the accuracy and the explainability of the predictions.
 
It is worth noting that the neighborhood size used to compute the explainability scores acts like a control parameter that could lead to a trade-off between accuracy and explainability: Notice how more neighbors increases the explainability precision (MEP) but only slightly (and slowly) reduces MAP; however even with this small decrease, MAP for E-AutoRec remains higher than for the non-explainable baseline (AutoRec).
We also investigated the effect of the explainability threshold $\theta$ on the explainability metric(Figure~\ref{fig:6(c)}). While our proposed E-AutoRec model maintains its higher explainability in comparison with the conventional AutoRec, the explainability decreases for both methods as we impose more restrictions (higher $\theta$) on the explainability score. This result is expected since a higher threshold reduces the number of explainable items relative to a user (resulting in more zeros in the explainability matrix $\boldsymbol{E}$). 

\section{CONCLUSION}
We proposed an explainable Autoencoder approach (E-AutoRec) which incorporates user ratings into both the inputs to be reconstructed and user-neighborhood collaborative filtering justified explainability scores as side information. Our experiments showed that E-AutoRec outperforms the baseline in terms of RMSE, explainability and MAP metrics for a wide range of tested parameters. The results demonstrate that using the pre-computed explainability scores as an additional side information fed to the input layer helps the Autoencoder to make predictions that are simultaneously more accurate and explainable. Thus there was no sacrifice in accuracy to pay for the increase in explainability of E-AutoRec. This is because the explainability values are integrated within the actual machine learning process, which keeps the same reconstruction loss function as the pure AutoRec model. However explainability is enhanced specifically because the explainabilty scores are used as side information to compensate for sparse inputs in the Autoencoder architecture, thus effectively enhancing both the accuracy and the explainability of the predictions. In the future, we plan to compare the proposed model with additional baseline models and investigate additional explainable Deep Learning architectures. 

%
\bibliographystyle{ACM-Reference-Format}
\bibliography{sample-base}

\end{document}